\documentclass{article}

\newtheorem{theorem}{Theorem}
\newtheorem{corollary}{Corollary}
\newtheorem{lemma}{Lemma}
\newtheorem{definition}{Definition}
\newcommand{\blackslug}{\mbox{\hskip 1pt \vrule width 4pt height 8pt 
depth 1.5pt \hskip 1pt}}
\newcommand{\QED}{\quad\blackslug\lower 8.5pt\null\par\noindent}

\newcommand{\cH}{\mbox{${\cal H}$}}

\newcommand{\cR}{\mbox{${\cal R}$}}

\newcommand{\eqdef}{\stackrel{\rm def}{=}}

\begin{document}

\title{Expected Qualitative Utility Maximization\thanks{This work was partially 
supported by the Jean and Helene Alfassa fund for 
research in Artificial Intelligence and by grant 136/94-1 of the 
Israel Science Foundation on ``New Perspectives on Nonmonotonic Reasoning''.}
}

\author{Daniel Lehmann\\Institute of Computer Science, Hebrew University\\Jerusalem 91904, Israel\\lehmann@cs.huji.ac.il}
\date{}
\maketitle

\abstract{A model for decision making that generalizes Expected Utility
Maximization is presented. This model, Expected Qualitative Utility 
Maximization, encompasses the Maximin criterion.
It relaxes both the Independence and the Continuity postulates.
Its main ingredient is the definition of a qualitative order
on nonstandard models of the real numbers and the consideration
of nonstandard utilities.
Expected Qualitative Utility Maximization
is characterized by an original weakening of von Neumann-Morgenstern's
postulates.
Subjective probabilities may be defined from those weakened postulates, 
as Anscombe and Aumann did from the original postulates.
Subjective probabilities are numbers, not matrices as in the Subjective
Expected Lexicographic Utility approach.
JEL no.: D81 Keywords: Utility Theory, Non-Standard Utilities, 
Qualitative Decision Theory}

\section{Introduction}
\label{sec:intro}
This paper proposes a deviation from the von Neumann-Morgenstern (vNM)
postulates
paradigm in decision theory.
It suggests that it may, in certain circumstances, be rational to judge
alternatives according to the issues in their main focus.
If these issues are deemed as equal, the
alternatives will continue to be judged as equal even if they are mixed with
different side-effects - side-effects which would not be judged as equal
were they to be in the main focus themselves. This contradicts the vNM
Independence axiom. Moreover, outcomes of issues in the main focus may
completely overshadow the outcomes of the side-issues, so alternatives of
the former may be judged as infinitely more (or infinitely less) preferable
than those of the latter. This contradicts the vNM Continuity axiom.

Suppose you have to choose between two lotteries.
In the first lottery you may win, with probability $p$,
a week's vacation in Hawaii. 
With probability $1 - p$ you get nothing.
In the second lottery you may win, with the {\em same} probability $p$,
the same vacation in Hawaii, but, with probability $1 - p$ you get
a consolation prize: a free copy of your favorite magazine.
Since the free copy is preferred to nothing, vNM's
independence postulate implies that lottery two is preferred to lottery one.
But couldn't a rational decision maker be indifferent
between the two lotteries.
One, often, I think, buys a lottery ticket in a frame of mind
focused on the big prize and not on the consolation prize.
This behavior is by no means general, as attested by the fact 
lotteries often
offer consolation prizes, but should a decision maker indifferent
between the two lotteries be considered {\em irrational}
in all situations?
I think not.

A variation on this example considers also a third lottery,
in which one wins a trip to Paris with the same probability $p$
as above, and nothing with probability $1 - p$.
Suppose you try to compare lotteries one and three.
You ponder at length the advantages and disadvantages of the
two vacation spots, and decide you are indifferent between the trip to Hawaii
and the trip to Paris, all relevant considerations taken into
account.
You conclude that you are also indifferent between the lotteries one and
three.
The independence axiom of von Neumann and Morgenstern
implies that lottery two is preferred to lottery three.
But it has been argued that it is quite unreasonable to expect
the very slight improvement that lottery two presents over lottery
one to overcome the lengthy and delicate deliberation that made
you conclude that the trips to Hawaii and Paris are equivalent for you.

Similar examples have been put forward to argue that the indifference 
relation is not always transitive:
lottery three is equivalent to lottery one and to lottery two,
but lottery two is preferred to lottery one.
The system presented in this paper endorses the transitivity
of indifference, but allows a decision maker to be indifferent
between lotteries one and two.

Let us consider now a second situation.
A patient has to choose between two options:
\begin{enumerate}
\item \label{one} (option $p$) do nothing and die in a matter of weeks,
\item \label{two} (option $q$) undergo surgery, the result of which,
depending on some objective probabilities, may be a long and happy life
(denoted by $l$)
or an immediate death on the operation table (denoted by $d$). 
We shall denote by $\lambda$ the
probability of the surgery being successful, i.e., the probability
of $l$. 
The probability of death on the operation table is, therefore, $1 - \lambda$.
\end{enumerate}
In other terms, one has to compare {\em mixtures} $p$ and
\mbox{$\lambda \, l + (1 - \lambda) \, d$}.
Assuming one prefers $l$ to $p$ and prefers $p$ to $d$,
vNM's postulates imply what we shall
call property {\bf P}: 
there is a unique \mbox{$\lambda \in ]0 , 1[$} for which 
one is indifferent between $p$ and
\mbox{$\lambda \, l + (1 - \lambda) \, d$}.
For any \mbox{$\mu > \lambda$}, one prefers 
\mbox{$\mu \, l + (1 - \mu) \, d$} to $p$, and for all 
\mbox{$\mu < \lambda$}, one prefers $p$ to \mbox{$\mu \, l + (1 - \mu) \, d$}.
If one thinks that a long and happy life is overwhelmingly
preferable to $p$ so as to make the distinction between $p$ and $d$
insignificant, i.e., that
\mbox{$\mu \, l + (1 - \mu) \, d$} is preferable to $p$ 
for any \mbox{$\mu \in ]0 , 1[$},
property {\bf P} fails and 
one deviates from vNM's rationality postulates.

But would we really dismiss as {\em irrational} such a behavior,
or such a preference?
After considering the question, in this section, I will propose
a weakening of vNM's postulates that allows
for the preference of \mbox{$\mu \, l + (1 - \mu) \, d$} over $p$
for any \mbox{$\mu \in ]0 , 1[$}.

Von Neumann-Morgenstern's point of view is perfectly acceptable
and I have no criticism for someone who adheres to it and decides
there is indeed some $\mu$, very close to zero, perhaps, such that
$p$ is equivalent to \mbox{$\mu \, l + (1 - \mu) \, d$}.
My only claim is that someone who thinks 
\mbox{$\mu \, l + (1 - \mu) \, d$} is preferable to $p$ 
for any \mbox{$\mu \in ]0 , 1[$}
cannot be considered {\em irrational} outright.

An argument, very similar to the one just presented,
for preferring a mixture \mbox{$\mu \, l + (1 - \mu) \, d$} to
some $p$, for any \mbox{$\mu \in ]0 , 1[$}, even though
$p$ is preferred to $d$, appears in Pascal's~\cite{Pascal:Pensees}.
There $l$ denotes eternal bliss (the reward of the believer
if God exists), $d$ denotes a life spent in error by a believer
in a God that does not exist, and $p$ denotes a life spent
by a non-believer.
This argument, known as Pascal's wager,
is very well-known and the reader may find in a previous version of this 
work~\cite{Util:Tech2}, a fuller discussion.

Before entering a technical description, I will, 
in short, describe intuitively the reason for the failure of property {\bf P}
in the system presented in this work, Expected Qualitative Utility 
Maximization.
In the examples above one may assume that
consequence $l$ is {\em infinitely} preferable
to $p$ and $d$.
This will be technically translated in giving an infinite value to
the utility of $l$, whereas $p$ and $d$ have finite utilities.
For von Neumann and Morgenstern, there are no infinite utilities.
This work develops a framework in which there are infinite utilities, 
by using nonstandard (in the sense
of A. Robinson~\cite{Robinson:66}) utilities.

A number of papers~\cite{Brito:75,Shapley:77,Aumann:77} discussed,
in the setting of the St. Petersburg's paradox, the existence of unbounded
utilities.
In the last of these papers, Aumann argues very convincingly
that utilities should be bounded.
At first sight, one may think that infinite utilities imply
unbounded utilities, and therefore Aumann argues also
against infinite utilities, but this is not the case.
His argument may be summarized in the following way:
if utilities were unbounded, for any \mbox{$\lambda \in ]0 , 1[$} 
there would be a consequence ($c$)
such that a lottery 
\mbox{$\lambda \, c + (1 - \lambda) \, d$}
is preferred to a long and happy life ($l$).
But this seems very unreasonable to Aumann.
His argument is directed against {\em unbounded} utilities,
but ineffective against {\em infinite} utilities.
Certainly, no consequence is infinitely preferable to $l$ and therefore,
if there are infinite utilities, the utility of $l$ is infinite.
But there is absolutely no problem if one assumes that
the utility of $l$ is infinite and maybe even maximal 
(nothing is preferred to $l$). In this case,
Aumann's argument disappears. The fact that $l$ is infinitely preferred
to some other consequence, $d$ for example, or a sum of money,
will influence the preferences of a decider between lotteries involving
$l$ and consequences such as $d$.
\section{Background}
\label{sec:back}
Utility theory is discussed in the framework 
of~\cite[Chapter 3]{vonNeuMorg:47}, see also~\cite{Fish:handbook}.
Let $\cH$ be a boolean algebra of subsets of $X$, and $P$ a convex set
of probability measures on $\cH$.
Convex means here that: 
\[
\forall p , q \in P , \: \forall \lambda \in ]0 , 1[, 
\ \lambda p + (1 - \lambda) q \in P.
\]
Here \mbox{$]0 , 1[$} denotes the standard open real interval.
Von Neumann and Morgenstern have characterized the binary relations
$>$ on $P$ that can be defined by a linear functional $u$ on $P$ in 
the following way:
\begin{equation}
\label{eq:fund}
\forall p , q \in P , \ p > q \: \Leftrightarrow \: u(p) > u(q).
\end{equation}
In Equation~\ref{eq:fund}, the functional $u$ is a function from $P$ to
the standard set of real numbers ${\bf R}$ and the relation $>$ in the
right hand side is the usual strict ordering on ${\bf R}$.

Their characterization is equivalent to the following, 
due to Jensen~\cite{Jen:67} 
(see~\cite[p. 1408]{Fish:handbook})
three conditions, for all \mbox{$p , q , r \in P$} and all 
\mbox{$\lambda \in ]0 , 1[$} 
(a weak order is an asymmetric and negatively transitive binary relation):
\[
{\bf A1} \ \ > {\rm \ on \ } P {\rm \ is \ a \ weak \ order} , 
\]
\[
{\bf A2} \ \ \ p > q \: \Rightarrow \: \lambda p + (1 - \lambda) r > 
\lambda q + (1 - \lambda) r,
\]
\[
{\bf A3} \ \ \ (p > q , q > r) \: \Rightarrow \: \exists \alpha , \beta \in ]0 , 1[ ,
{\rm \ such \ that \ } 
\]
\[
\ \ \ \ \ \ \ \ \ \alpha p + (1 - \alpha) r > q > \beta p + (1 - \beta) r.
\]
The three conditions above are not the original postulates of
von Neumann and Morgenstern, they are equivalent to them.
They will be referred to, nevertheless, in this work, as 
vNM's postulates.
{\bf A2} is usually denoted Independence and {\bf A3} Continuity.
The purpose of this work is to generalize von Neumann and Morgenstern's
characterization to deal with {\em qualitative} utilities.
In the sequel, \mbox{$p \geq q$} will denote \mbox{$q \not > p$} and
\mbox{$p \sim q$} will denote the conjunction of \mbox{$p \geq q$} and
\mbox{$q \geq p$}.
\section{Qualitative utilities}
Qualitative decision theory has been developed mostly in opposition to
quantitative decision theory, stressing decision methods that do not 
satisfy vNM's or Savage's~\cite{Savage:54} postulates, 
the postulates generally accepted for quantitative decision theory.
The focus in qualitative decision theory has always been on methods
and algorithms, more than on an axiomatic treatment (\cite{BrafTen:qdt}
is an exception).

A different approach is proposed here: qualitative and quantitative
decision theories can be viewed as special cases of a unified general
theory of decision that contains both.
This unified theory is a generalization of the quantitative theory.
The power of the generalization lies in the consideration of nonstandard
models of the set of real numbers for utilities
and a definition of indifference that neglects infinitesimally small
differences.
In this paper probabilities will always be standard.
The case of nonstandard probabilities and standard utilities has been
treated, after the elaboration of this paper, in~\cite{LehNonst:TARK}.
Preliminary ideas already appeared in~\cite{Leh:UAI96} and~\cite{Leh:QDT}.

A well-established tradition in Decision Theory considers Expected Utility
Maximization as the only rational policy.
Following this view, an act $f$ is strictly preferred to an act $g$
iff the utility expected from $f$ is strictly larger than that
expected for $g$.
Since expected utilities are real numbers, {\em strictly larger} has
its usual, {\em quantitative} meaning.
The main claim of this paper is that the qualitative point of view
may be subsumed by a slightly different definition of {\em strictly larger}.
Suppose we consider any model elementarily equivalent to the real numbers,
more precisely, any (standard or nonstandard) model of the real numbers,
$\cR$ (for the standard model, we shall use ${\bf R}$).
Let $x$ and $y$ be elements of $\cR$. 
To make matters simpler, suppose that both $x$ and $y$ are positive.
The number $x$ is quantitatively
larger than $y$ iff \mbox{$x - y > 0$}.
What could be a reasonable definition of {\em qualitatively larger}?
Clearly, if $x$ is qualitatively larger than $y$ then it must be quantitatively
larger: in a sense {\em qualitatively larger} means {\em definitely larger}.
A first idea that may be considered is to use a notion that proved
fundamental for nonstandard analysis (the monads of~\cite{Robinson:66},
or see~\cite{Keisler:76}):
the notion of two numbers being {\em infinitely close}, 
and consider that a number
$x$ is {\em qualitatively}
larger than a number $y$ iff $x$ is larger than $y$ and {\em not
infinitely close} to $y$, i.e., iff \mbox{$x - y$} is 
strictly larger than some positive standard number.
At first this idea looks appealing:
if $\epsilon$ is strictly positive and infinitesimally close to zero,
and $x$ is a standard, strictly positive, real,
we do {\em not} want \mbox{$x + \epsilon$} to be qualitatively larger than
$x$.
At a second look, one realizes that the size of \mbox{$x - y$} should not
be judged absolutely, but relatively to the size of $x$: for example
\mbox{$\epsilon^{2} + \epsilon$} should be qualitatively larger than
$\epsilon^{2}$. Therefore I propose the following definition:
\begin{definition}
\label{def:qual>}
Let $x$ and $y$ be positive. We shall say that $x$ is 
qualitatively larger than $y$ and write \mbox{$x \succ y$} iff 
\mbox{${x - y} \over {x}$} is strictly positive and {\em not} infinitesimally
close to zero:
in other terms, iff there is a strictly positive {\em standard} number
$r$ such that \mbox{${{x - y} \over {x}} \geq r$}.
\end{definition}
To the best of my knowledge, the idea of comparing nonstandard numbers
qualitatively, i.e., by $\succ$ and not by the usual ordering $>$ is original.
Notice also that Definition~\ref{def:qual>} relies on the notion
of a {\em nonstandard} number, and that this notion 
is not first-order definable. We may therefore expect for the qualitative
order properties that are quite different than the ones enjoyed by the
usual order. The qualitative order is {\em not} a lexicographic ordering.
It will be shown in Section~\ref{sec:previous} that its properties
are very different from those of lexicographic orderings.

The definition may be extended to arbitrary numbers in the obvious way:
\begin{enumerate}
\item
if \mbox{$x \geq 0$} and \mbox{$0 > y$}, then \mbox{$x \succ y$}, and
\item
\mbox{$x \succ y$} iff \mbox{$-y \succ -x$},
\end{enumerate}
but we shall limit ourselves to the consideration of positive utilities
in this paper.
\mbox{$x \preceq y$} shall denote \mbox{$ x \not \succ y$} 
and \mbox{$x \sim y$} shall denote that 
\mbox{$x \preceq y$} and \mbox{$y \preceq x$}.
Properties of the qualitative ordering $\succ$ are described and proved
in Appendix A.

Note that \mbox{$x \preceq y$} does not imply
\mbox{$x - c \preceq y - c$}.
If $\epsilon$ is positive and infinitesimally close to zero,
\mbox{$1 + \epsilon \preceq 1$} but 
\mbox{$1 + \epsilon - ( 1 - \epsilon^{2}) \not \preceq \epsilon^{2}$}.
Note also that $\succ$ is not complete and that \mbox{$\preceq$} is not
antisymmetric.
Notice that, if we choose, for $\cR$, the standard model of the reals, 
${\bf R}$,
then \mbox{$x \succ y$} iff \mbox{$x > y$}.
Therefore our treatment would include the classical approach, 
if we allowed also negative utilities. As said above, in this paper,
we concentrate on the case all utilities are positive.
Is our framework, with positive nonstandard utilities, a generalization
of the classical theory, with standard positive and negative utilities?
Since, in the classical setting, utilities are defined only up
to an additive constant, bounded utilities may always be
considered to be positive, 
by adding a positive large enough constant.
In view of Aumann's~\cite{Aumann:77} critique of unbounded utilities,
we feel that the present framework encompasses the most important
part of classical theory.

Expected Qualitative Utility Maximization is the version of
Expected Utility Maximization that obtains when, for utilities, 
\begin{itemize}
\item the model chosen for the real numbers {\em may} be nonstandard, and
\item real numbers are compared {\em qualitatively}, i.e., by $\succ$.
\end{itemize}
We want to characterize the binary relations
$>$ on $P$ that can be defined by a linear functional 
\mbox{$u: P \rightarrow \cR_{+}$} in 
the following way:
\begin{equation}
\label{eq:nonst}
\ \ \ \ \forall p , q \in P , p > q \Leftrightarrow u(p) \succ u(q).
\end{equation}
One could, as well, consider functionals of the type:
\mbox{$u: P \rightarrow \cR$}, but the characterization of relations
obtained in such a general way is more intricate than the task
proposed in this paper and is left for future work.
We shall note the main difference between the two characterizations
in the sequel.
One should immediately notice that, by Lemma~\ref{le:basic} part~\ref{cx},
if \mbox{$c > 0$}, the utility function $c \: u(p)$ defines the same
ordering as $u(p)$.
But, contrary to what happens both in the classical and
in the lexicographic ordering settings, if $d \in \cR$
the function \mbox{$d + u(x)$} does not, in general, define the same
ordering as $u(p)$: consider Lemma~\ref{le:basic} part~\ref{c+2}.
Such an instability under an additive constant, and in particular
an asymmetry between gains and losses has
been found in the behavior of decision makers 
in many instances~\cite{FishKoch:79,KahnTver:79,Schoe:80}.
The question of whether Expected Qualitative Utility Maximization,
in the present form, or in the generalized version that allows for negative 
utilities,
is a realistic model for explaining such behavior cannot be discussed
in this paper.
\section{Maximin as Expected Qualitative Utility Maximization}
As noticed above, Expected Qualitative Utility Maximization generalizes
Expected Utility Maximization: if one chooses the standard model for
real numbers then Expected Qualitative Utility Maximization 
boils down exactly to Expected Utility Maximization, at least when
utilities are bounded.
We shall show now that considering nonstandard utilities enables us
to obtain decision criteria that do not satisfy 
vNM's postulates and were so far considered as
part of the realm of {\em qualitative} decision theory.
A version of the Maximin criterion will be presented.
The Maximin criterion has been proposed by A. Wald~\cite{Wald:50},
in a different framework. The criterion to be presented is
a variation on this theme.

Notice that any bona fide Maximin criterion violates Independence,
and is therefore not obtainable by a lexicographic ordering, since all
such methods satisfy Independence.
Suppose \mbox{$p \succ q \succ r$}. Then, under the maximin criterion 
\mbox{$\lambda r + (1 - \lambda) p$} must be considered indifferent to
\mbox{$\lambda r + (1 - \lambda) q$} since in both the worst possible outcome
is the same and obtains with the same probability, violating Independence.

Assume the set $X$ is finite and $\cH$ contains all subsets of $X$.
Let the elements of $X$ be \mbox{$x_{0}, \ldots, x_{n - 1}$}.
Let $\epsilon$ be a number that is positive and infinitesimally close to zero
and let our utility function $u$ be the linear function defined by:
\mbox{$u(x_{i}) = - \epsilon ^ {n - i - 1}$}, 
for \mbox{$i = 0 , \ldots , n - 1$}. Notice that, exceptionally, 
we consider here negative utilities.
Notice that \mbox{$x_{i} < x_{j}$} iff \mbox{$i < j$}.
The utility of a mixture 
\mbox{$\lambda x_{i} + (1 - \lambda) x_{j}$} is
\mbox{$- \lambda \epsilon ^ {n - i - 1}$} if \mbox{$x_{i} < x_{j}$} and
\mbox{$- \epsilon^{n - i - 1}$} if \mbox{$x_{i} \sim x_{j}$}.
Suppose \mbox{$i < j$} and \mbox{$i ' < j '$}.
Then, 
\mbox{$\lambda x_{i} + (1 - \lambda) x_{j} <$}
\mbox{$\mu x_{i '} + 
(1 - \mu) x_{j '}$} holds iff \mbox{$x_{i} < x_{i '}$}
or \mbox{$x_{i} \sim x_{i '}$} and \mbox{$\lambda > \mu$}.
The decision maker therefore compares different mixtures by comparing
the worst possible outcomes and, if they are the same, their respective
probabilities. This is some form of Maximin criterion
and does not satisfy {\bf A2} or {\bf A3}, but it has been considered
a rational way of deciding by many authors, and it
is amenable to Expected Qualitative Utility Maximization.
\section{Postulates}
\label{sec:post}
Suppose $>$ is defined by some functional $u$ as in Equation~\ref{eq:nonst}.
First, one easily sees that the relation $>$ 
defined by Equation~\ref{eq:nonst} 
is a weak order since $\succ$ is a weak order.
Therefore {\bf A1} holds.

Let us know consider {\bf A2}.
We shall see that {\bf A2} does not hold.
Let $\epsilon$ be positive and infinitesimally close to zero.
Let \mbox{$u(p) = 2 \times \epsilon$}, \mbox{$u(q) = \epsilon$}
and \mbox{$u(r) = 1$}.
Then \mbox{$0.5 \times 2 \times \epsilon + 0.5 \sim 0.5 \times \epsilon + 0.5$}
and \mbox{$0.5 p + 0.5 r \sim 0.5 q + 0.5 r$}.
Let us find a weakening of {\bf A2} that holds.
Assume now that \mbox{$p > q$}, i.e., \mbox{$u(p) \succ u(q)$}. 
Let \mbox{$\lambda \in {\bf R}$}, be a standard 
real number, \mbox{$0 < \lambda < 1$}.
Let us compare \mbox{$\lambda p + (1 - \lambda) r$}
and \mbox{$\lambda q + (1 - \lambda) r$}.
Since $u$ is linear, 
\[ 
u(\lambda p + (1 - \lambda) r) = \lambda u(p) + (1 - \lambda) u(r)
\]
and
\[
u(\lambda q + (1 - \lambda) r) = \lambda u(q) + (1 - \lambda) u(r).
\]
By Lemma~\ref{le:basic}, parts~\ref{c+1} and~\ref{c+2}, 
two cases must be considered.
If \mbox{$(1 - \lambda) u(r) / \lambda u(p)$} is finite,
i.e., if \mbox{$u(r) / u(p) $} is finite, then,
\[
\lambda u(p) + (1 - \lambda) u(r) \succ \lambda u(q) + (1 - \lambda) u(r),
\]
and therefore 
\mbox{$\lambda p + (1 - \lambda) r > \lambda q + (1 - \lambda) r$},
which corresponds exactly to {\bf A2} above.
If, on the contrary, \mbox{$u(r) / u(p)$} is infinite, then, 
for any \mbox{$\lambda' \in ]0 , 1[$} (the {\em standard} unit interval),
\[
(1 - \lambda') u(r) \sim \lambda' u(p) + (1 - \lambda') u(r)
\sim \lambda' u(q) + (1 - \lambda') u(r) \succ u(p).
\]

To formulate our independence property, it is best to set 
the following definition.
\begin{definition}
\label{def:>>}
We shall say that $p$ overrides $q$ and write \mbox{$p \gg q$} iff
\mbox{$p > q$} and for any $q'$ such that \mbox{$q > q'$}
and for any \mbox{$\lambda \in ]0 , 1[$},
\mbox{$\lambda q + (1 - \lambda) p \sim$}
\mbox{$\lambda q' + (1 - \lambda) p$}.
\end{definition}
The intuitive meaning of \mbox{$p \gg q$} is that $p$ is so much preferred
to $q$ that, in any lottery in which $p$ and $q$ are the
prizes, if one does not win $p$, one does not even care to cash $q$, 
but would as well get any lesser prize $q'$.
Notice that, since $>$ is asymmetric, the relation $\gg$ is also
asymmetric and therefore irreflexive.
The following lemma describes the tight relationship between the properties
\mbox{$p \gg q$} and \mbox{$u(p) / u(q)$} is infinite.
\begin{lemma}
\label{le:inf}
If the relation $>$ is defined by a linear functional
as in Equation~\ref{eq:nonst}, then 
\mbox{$p \gg q$} iff either \mbox{$ u(p) / u(q) $} is infinite,
or \mbox{$p > q$} and 
\mbox{$q \leq r$} for any \mbox{$r \in P$}.
\end{lemma}
\begin{proof}
Let us proof the {\em if} part.
If \mbox{$q \leq r$} for any \mbox{$r \in P$} and \mbox{$p > q$}, 
then obviously
\mbox{$p \gg q$}.
If \mbox{$ u(p) / u(q) $} is infinite, then \mbox{$u(p) \succ u(q)$}
and, if \mbox{$u(q) > u(q')$}, 
\[
\lambda u(q) + (1 - \lambda) u(p) \sim
(1 - \lambda) u(p) \sim
\lambda u(q') + (1 - \lambda) u(p).
\]

Let us prove now the {\em only if} part.
If \mbox{$u(q) \succ u(q')$} and
\[
\lambda u(q) + (1 - \lambda) u(p) \sim
\lambda u(q') + (1 - \lambda) u(p),
\]
then 
\mbox{$\lambda (u(q) - u(q')) / (\lambda u(q) + (1 - \lambda) u(p)$}
is infinitesimally close to zero and therefore
\[
\frac{u(q)}{u(q) - u(q')} + \frac{1 - \lambda}{\lambda} \ 
\frac{u(p)}{u(q) - u(q')}
\]
is infinite.
But \mbox{$u(q) \succ u(q')$} implies the first term is not infinite.
Therefore the second one is.
Then \mbox{$u(p) / (u(q) - u(q'))$} is infinite.
But \mbox{$u(q) - u(q') >$}
\mbox{$\gamma u(q)$}, for some {\em standard}
positive $\gamma$, 
\mbox{$u(p) / (u(q) - u(q')) <$}
\mbox{$u(p) / \gamma u(q)$} and
\mbox{$u(p) / u(q)$} is infinite.
\end{proof}
Our independence property may now be formulated as:
\[
{\bf A'2} \ \ p > q , r \not \gg p \: \Rightarrow \: 
\forall \lambda \in ]0,1[  
\ \lambda p + (1 - \lambda) r > \lambda q + (1 - \lambda) r.
\]
The intuitive meaning of {\bf A'2} is that any lottery is sensitive to
both its prizes, unless one of the prizes overrides the other one.

Let us study now property {\bf A3}.
Let us show it does not hold either.
Let $\epsilon$ be positive and infinitesimally close to zero.
Let \mbox{$u(p) = 1$}, \mbox{$u(q) = 2 \epsilon$}, 
\mbox{$u(r) = \epsilon$}.
We have \mbox{$p > q > r$}, but for any
\mbox{$\beta \in ]0 , 1[$},
\mbox{$u(q) \leq$}
\mbox{$\beta u(p) + (1 - \beta) u(r)$}.

Let us find a sound weakening of {\bf A3}.
Assume \mbox{$p > q > r$}.
We have \mbox{$u(p)  \succ$}
\mbox{$u(q) \succ$} \mbox{$u(r)$}.
Consider the values \mbox{$\alpha u(p) + (1 - \alpha) u(r)$}
for the different values of \mbox{$\alpha \in ]0 , 1[$}.
Since \mbox{$u(p) \succ u(q)$}, there is some standard \mbox{$\gamma > 0$}
such that \mbox{$(u(p) - u(q)) / u(p) >$} \mbox{$\gamma$}.
We have \mbox{$(1 - \gamma) u(p) >$} \mbox{$u(q)$}.
Let
\mbox{$\alpha = 1 - \gamma$}, \mbox{$\alpha \in ]0 , 1[$}.
We have 
\[
\alpha u(p) + (1 - \alpha) u(r) = (1 - \gamma) u(p) + \gamma u(r) > 
u(q) + \gamma u(r) > u(q).
\]
Take any \mbox{$\alpha' > \alpha$}, \mbox{$\alpha' < 1$}.
We have
\mbox{$\alpha' u(p) + (1 - \alpha') u(r) \succ$} \mbox{$u(q)$}.
We have shown that \mbox{$\alpha' p + (1 - \alpha') r >$} \mbox{$q$}.

Let us now consider the existence of a \mbox{$\beta \in ]0 , 1[$}
such that 
\mbox{$q > \beta p + (1 - \beta) r$}.
We have \mbox{$u(q) / u(p) \succ$} \mbox{$u(r) / u(p)$}.
If \mbox{$u(q) / u(p)$} is not infinitesimal,
there is some standard 
\mbox{$\gamma < 1$} such that \mbox{$u(q) / u(p) \succ$}
\mbox{$\gamma \succ$}
\mbox{$u(r) / u(p)$}.
If \mbox{$\beta > 0$} approaches zero, 
\mbox{$(\gamma - \beta) / (1 - \beta)$} approaches $\gamma$ from 
below. There is therefore some standard \mbox{$\beta \in ]0 , 1[$} 
such that
\mbox{$\gamma \succ (\gamma - \beta) / (1 - \beta) \succ$}
\mbox{$u(r) / u(p)$}.
Therefore, we have \mbox{$(\gamma - \beta) u(p) \succ$}
\mbox{$(1 - \beta) u(r)$}, and
\mbox{$\gamma u(p) \succ$} \mbox{$\beta) u(p) + (1 - \beta) u(r)$}.
But \mbox{$u(q) \succ$} \mbox{$\gamma u(p)$}.
We conclude that \mbox{$ q >$} \mbox{$\beta p + (1 - \beta) r$}.
If, on the contrary, \mbox{$u(q) / u(p)$} is infinitesimal,
for any \mbox{$\beta \in ]0 , 1[$}, \mbox{$\beta u(p) \succ u(q)$}
and \mbox{$\beta p + (1 - \beta) r >$} \mbox{$q$}.
In this case, note that, for any \mbox{$\beta \in ]0 , 1[$},
\[
\beta u(p) + (1 - \beta) u(r) \sim  \beta u(p) \sim 
\beta u(p) + (1 - \beta) u(q).
\]

We conclude that the following properties hold:
\[
{\bf A'3} \ \ p > q > r \: \Rightarrow \: 
\exists \alpha \in ]0 , 1[ {\rm \ such \ that \ }
\alpha p + (1 - \alpha) r > q .
\]
\[
{\bf A''3} \ \ p > q > r , p \not \gg q \: \Rightarrow \: 
\exists \beta \in ]0 , 1[ {\rm \ such \ that \ }
q > \beta p + (1 - \beta) r.
\]
\section{Representation Result}
All the technical work needed to prove our theorem may be
found in Appendix B.
The general structure of the proof is very similar to that 
of the classical proof of the completeness of Jensen's postulates
for representation by a standard linear functional. Nevertheless,
there are a few subtle points, since one has to carefully avoid using
those properties that hold in the standard case but do not hold here.
The main result of this work may now be presented.
\begin{theorem}
\label{the:main}
If the set $P$ is finitely generated, then the relation
$>$ satisfies {\bf A1}, {\bf A'2}, {\bf A'3} and {\bf A''3}
iff there is some nonstandard model \cR\ and some
linear functional \mbox{$u : P \longrightarrow \cR_{+}$} such that
\mbox{$\forall p , q \in P$}, \mbox{$p > q$} iff
\mbox{$u(p) \succ u(q)$}.
\end{theorem}

The proof of Theorem~\ref{the:main} is contained in 
Appendix B.
One may certainly weaken the restriction to a finitely generated
$P$ a bit, but some restriction is needed here.
The following example will show that Theorem~\ref{the:main} does not
hold if one removes the condition that $P$ be finitely generated. 
Consider, for example, any nonstandard model of the real
numbers and let $P$ be the subset
including all positive numbers that are neither infinite nor 
infinitesimally close to zero.
Let us consider the usual ordering $>$ ({\em not} $\succ$) on $P$.  
This ordering is easily seen to satisfy 
vNM's postulates.
Suppose there is a nonstandard model \cR\ and a linear
function \mbox{$u : P \longrightarrow \cR_{+}$} such that
\mbox{$\forall p , q \in P$}, \mbox{$p > q$} iff
\mbox{$u(p) \succ u(q)$}.
Since postulate {\bf A2} (independence) holds,
the image of $u$ has to be contained in one single
equivalence class under $\asymp$.
All such equivalence classes are isomorphic (by multiplication).
We may therefore assume that the image of $u$ is contained in the class
of numbers that are neither infinite nor infinitesimally close to zero.
Any member of this class is $\sim$-equivalent to a unique standard
real.
We may therefore take $u$ into the standard real numbers.
But this is impossible if the nonstandard model chosen has a cardinality 
that is larger than the continuum. In this case, $P$ has 
the same cardinality and, since $u$ must be one-to-one,
its image has a cardinality larger than that of ${\bf R}$.
\section{Comparison with previous work}
\label{sec:previous}
Numerous works during the fifties and the sixties considered
weakenings of vNM's postulates,
see~\cite{Fish:handbook} for a survey.
Some of these weakenings try to describe the actual preferences
of human deciders and reject the Independence property {\bf A2}.
Often they also reject the requirement that preferences be transitive,
or that indifference be transitive.
Others, based on lexicographic orderings, keep the Independence
postulate and reject only the Continuity postulate {\bf A3}:
\cite{Hausner:54,Chipman:60,Chipman:71,Fishburn:lexi71,Fishburn:lexi74,
BlumeBranDek:91,LaValleFish:91,LaValleFish:92, Rajan:Trembles}.
Nonstandard analysis~\cite{Robinson:66} appeared late on the scene.
Few authors applied it to decision theory: \cite{Skala:74,Kannai:92}.

This work proposes an original weakening based on
nonstandard analysis. We shall describe its relation to
lexicographic orderings first, and then to other uses of nonstandard
analysis in decision theory.

The motivation for this work is, in fact, opposite to the basic motivation
for considering lexicographic orderings.
The motivation for the consideration of lexicographic orderings on
probabilities is the consideration of those perfection refinements 
of Nash Equilibria which required optimization also off the
equilibrium path, on zero-probability events. Replacing the zero
probabilities by infinitesimal probabilities and requiring optimization in
the usual sense (i.e. with the usual linear ordering on (non-standard) numbers)
gives the desired result. The current paper is trying to achieve exactly the
opposite - ``side issues'' should not matter, and hence the special ordering
proposed. 

For a more technical comparison, notice that
our postulates are very close to the original postulates
of von Neumann and Morgenstern. In particular
the ordering $<$ is modular (weak total) 
and the indifference relation
$\sim$ is transitive. But both {\bf A2} and {\bf A3} are weakened,
in a closely linked manner.
Notice that lexicographic orderings
define a preference relation that satisfies {\bf A2}, and that
{\bf A2}, {\bf A'3} and {\bf A''3} together
imply {\bf A3}, since {\bf A2} says that \mbox{$p \gg q$}
implies that for any $w$, \mbox{$w \geq q$}.
The weakening offered by lexicographic orderings is 
therefore orthogonal to ours: any lexicographic ordering that satisfies 
our postulates satisfies the full set of
vNM's postulates.

To explain better the difference between lexicographic orderings
and our qualitative ordering, assume $P$ is the real plane ${\bf R}^{2}$
ordered by the lexicographic ordering: 
\[ (x , y) < (x' , y') \ {\rm iff \ either} \  x < x' 
\ {\rm or} \  x = x' \ {\rm and} \  y < y'.
\]
For \mbox{$\lambda \in [0 , 1]$}, define 
\mbox{$\lambda (x , y) + (1 - \lambda) (x' , y')$}
to be \mbox{$(\lambda x + (1 - \lambda) x' , \lambda y + (1 - \lambda) y')$}.
Notice that \mbox{$(0 , 0) <$} \mbox{$(1 , 10) <$} \mbox{$(2 , 0)$}, 
but there is no 
\mbox{$\lambda \in ]0 , 1[$} such that 
\mbox{$(1 , 10) =$} \mbox{$\lambda (0 , 0) + (1 - \lambda) (2 , 0)$}
since \mbox{$\lambda (0 , 0) + (1 - \lambda) (2 , 0) =$}
\mbox{$(2 (1 - \lambda) , 0)$}.
Both lexicographic and qualitative orderings imply the failure
of property {\bf P} of section~\ref{sec:intro}.
But notice that lexicographic ordering also implies the failure of
Lemma~\ref{le:exist0} since there exists some \mbox{$\beta \in ]0 , 1[$}
such that:
\[
(1 , 10) > \beta (2 , 0) + (1 - \beta) (0 , 0) =
\beta (2 , 0),
\]
for example \mbox{$\beta = 0.4$}, and nevertheless there is no $\lambda$
as above.
Lexicographic and qualitative orderings stem from different concerns
and have very different characteristics.

The previous uses of nonstandard numbers mentioned above consider
the usual ordering on those nonstandard models, {\em not} the qualitative
ordering we consider.
Most of the considerations of~\cite{Skala:74} concerning strategic games in 
which some pay-offs are infinite are valid in the present framework
too, and therefore very relevant for further work.
\section{Subjective probability}
\label{sec:Subj}
In~\cite{AnscAum:63}, Anscombe and Aumann showed that, in a slightly
modified framework in which the decider compares not lotteries but acts the 
outcomes of which depend on the state of nature,
an additional very natural postulate is sufficient
to ensure the existence of subjective probabilities, in the decider's
mind.
This result is described in Section~\ref{sub:AnscAum}.
In~\ref{sub:additional} and~\ref{sub:char}, it is shown that the same
holds, with minimal changes, for Expected Qualitative Utility 
Maximization.
This result must be contradistincted with 
LaValle and Fishburn's~\cite{LaValleFish:91}, showing that, in the
Lexicographic Subjective Expected Utility framework, subjective
probabilities have to be matrices rather than numbers.
\subsection{Anscombe-Aumann's framework}
\label{sub:AnscAum}
Let $P$ be a convex set of probability measures on $\cH$,
as in Section~\ref{sec:back}.
In~\cite{AnscAum:63}, Anscombe and Aumann consider
a finite set $S$ (of states) and the set ${\bf F}$ (of acts)
of mappings: \mbox{$S \longrightarrow P$}.
The set ${\bf F}$ is a mixture set when,
for \mbox{$a , b \in {\bf F}$}, \mbox{$\lambda \in [0 , 1]$},
one defines \mbox{$[\lambda a + (1 - \lambda) b](s)$} as
\mbox{$\lambda a(s) + (1 - \lambda) b(s)$}.
If $f$ is an element of $P$ we shall identify it with 
the constant function \mbox{$h \in {\bf F}$} defined by
\mbox{$h(s) = f$} for any \mbox{$s \in S$}.
They consider binary relations ($>$) on ${\bf F}$ that may be
defined a probability measure $p$ on $S$ and a linear function
\mbox{$u: P \longrightarrow {\bf R}$} by:
\[
a > b \Leftrightarrow 
\sum_{s \in S} p(s) u(a(s)) > \sum_{s \in S} p(s) u(b(s)).
\]
Assuming $P$ is finitely generated (in fact they have a slightly weaker
condition), 
they show that a relation $>$ on ${\bf F}$ may be defined by a
probability measure $p$ on $S$ and a linear $u$ in the way above
iff $>$ satisfies {\bf A1}, {\bf A2}, {\bf A3} (where $>$ is defined
on {\bf F}) and one 
additional postulate.
For any \mbox{$a , b \in {\bf F}$}
\[
{\bf A'4} \ {\rm \ If \ } \forall s \in S , s \neq s_{0} 
\Rightarrow a(s) = b(s), {\rm \ then \ }
a > b \: \Rightarrow \: a(s_{0}) > b(s_{0}).
\]
In the last part of {\bf A'4}, \mbox{$a(s_{0})$} and
\mbox{$b(s_{0})$} stand for the corresponding constant functions.
\subsection{Additional Postulate}
\label{sub:additional}
We shall need an additional postulate, specific to our consideration
of nonstandard utilities.
A definition will be handy.
Roughly, an element $t$ of $S$ is null iff the singleton $\{ t \}$
is Savage-null, i.e., if the value taken by an act on $t$
is insignificant.
\begin{definition}
\label{def:Lzero}
Let \mbox{$t \in S$}. The element $t$ is said to be {\em null}
iff for any \mbox{$a , b \in {\bf F}$} we have \mbox{$a \sim b$}
if $a$ and $b$ agree everywhere except possibly on $t$, i.e.,
for any \mbox{$s \neq t$}, \mbox{$a(s) = b(s)$}.
\end{definition}
Our last postulate ensures that probabilities are {\em standard}.
Assume \mbox{$t \in S$}, \mbox{$a \in {\bf F}$} and 
one of the possible values of $a$, $a(t)$, overrides $a$, 
i.e., \mbox{$a(t) \gg a$}.
It must be the case that $t$ is null.
\[
{\bf A'5} \ \ \ t \in S , \: a \in {\bf F} , \: a(t) \gg a \: \Rightarrow \:
t {\rm \ is \ null}.
\]
A first consequence of {\bf A'4} shall be proved now.
\begin{lemma}
\label{le:A'4}
Assume {\bf A1} and {\bf A'4} hold.
If \mbox{$a(s) \preceq b(s)$} for any \mbox{$s \in S$},
then \mbox{$a \preceq b$}.
\end{lemma}
\begin{proof}
By induction on the number of elements $s$ of $S$ 
for which \mbox{$a(s) \neq b(s)$}, using {\bf A'4}.
\end{proof}
\subsection{Characterization Theorem}
\label{sub:char}
A theorem similar to Anscombe and Aumann's holds for 
Expected Qualitative Utility Maximization.
\begin{theorem}
\label{the:subj}
Assume $P$ is finitely generated and $>$ is a binary relation on ${\bf F}$.
There is a probability measure $p$ on $S$, a nonstandard model $\cR$ 
and a linear function \mbox{$u : P \longrightarrow \cR_{+}$} such that
\mbox{$\forall a , b \in {\bf F}$}, 
\begin{equation}
a > b \Leftrightarrow
\sum_{s \in S} p(s) u(a(s)) \succ \sum_{s \in S} p(s) u(b(s))
\end{equation}
iff the relation $>$ satisfies {\bf A1}, {\bf A'2}, {\bf A'3}, 
{\bf A''3}, {\bf A'4} and {\bf A'5}.
If the relation $>$ on $P$ is not trivial, 
the probability measure $p$ is unique.
\end{theorem}
The proof may be found in Appendix C.
Its general structure is similar to Anscombe and Aumann's,
but there are a number of steps that have to be managed in a different
manner.
\section{Conclusion}
Nonstandard models of the real numbers provide for a natural notion of
qualitative equivalence and a principle of Qualitative Utility Maximization.
This work characterizes in full the situation in which one allows 
nonstandard utilities but insists on standard probabilities.
In this framework one may consider consequences that are infinitely
preferable to others and criteria of the Maximin family.
Subjective probabilities may be shown to exist as in the classical
framework. They are numbers, not matrices, as in the 
Lexicographic Subjective Expected Utility framework.
This represents probably a clear advantage for 
Qualitative Utility Maximization over lexicographic methods.
But both are really orthogonal extensions of the classical
Expected Utility Maximization paradigm.

The study of games with nonstandard utilities seems appealing.
The dual case of nonstandard probabilities and standard utilities
and the most general case of nonstandard probabilities and utilities
are yet to be characterized.
They will include consideration of subjective probabilities
infinitesimally close to zero.




\appendix
\section{Properties of the Qualitative Ordering}
The following lemma summarizes the main properties of the relations just
defined. Let $\cR_{+}$ be the subset of $\cR$ consisting of all
positive numbers (including zero).
\begin{lemma}
\label{le:basic}
For any \mbox{$x , y , z \in \cR_{+}$}
\begin{enumerate}
\item \label{<} \mbox{$x \succ y \Rightarrow x > y$},
\item \label{zero} \mbox{$x > 0 \Rightarrow x \succ 0$},
\item \label{asymmetric} the relation $\succ$ is asymmetric: i.e.,
\mbox{$x \succ y \Rightarrow y \preceq x$},
\item \label{succ>=} \mbox{$ x \succ y$} and \mbox{$y \geq z$}
imply \mbox{$x \succ z$},
\item \label{>=succ} \mbox{$ x \geq y$} and \mbox{$y \succ z$}
imply \mbox{$x \succ z$},
\item \label{trans} the relation $\succ$ is transitive,
\item \label{negtrans} the relation $\succ$ is negatively transitive, i.e., 
modular: i.e., 
\mbox{$x \preceq y$} and \mbox{$y \preceq z$} imply 
\mbox{$x \preceq z$},
\item \label{equ} the relation $\sim$ is an equivalence relation,
\item \label{cx} if \mbox{$c > 0$}, then \mbox{$x \succ y$} iff 
\mbox{$c x \succ c y$},
\item \label{c+1} if \mbox{$x \succ y$}, \mbox{$c > 0$} and
\mbox{$c / x$} is finite, then 
\mbox{$x + c \succ y + c$}.
\item \label{c+2}  if \mbox{$x \succ y$}, \mbox{$c > 0$} and
\mbox{$c / x$} is infinite, then, for any {\em standard} positive 
\mbox{$\lambda, \mu, \nu$},
\mbox{$\lambda c \sim$} \mbox{$\mu x + \lambda c \sim$}
\mbox{$\mu y + \lambda c \succ$} \mbox{$\nu x$},
\item \label{negc+} if \mbox{$x \succ y$}, \mbox{$c > 0$} and
\mbox{$y - c > 0$}, then 
\mbox{$x - c \succ y - c$}.
\end{enumerate}
\end{lemma}
\begin{proof}
Parts~\ref{<} and~\ref{zero} are obvious from the definition of $\succ$.

Part~\ref{asymmetric} follows from part~\ref{<}.

For part~\ref{succ>=}, 
notice that 
\mbox{$x - z \geq x - y$} and \mbox{$(x - z) / x \geq (x - y) / x$}.

For part~\ref{>=succ}, 
note that \mbox{$x - z \geq x - y$} and \mbox{$(x - z) / x \geq (x - y) / x$}.

Part~\ref{trans} follows from parts~\ref{<}, \ref{succ>=} and~\ref{>=succ}.

For part~\ref{negtrans}, assume \mbox{$x \succ y$}.
We shall show that either \mbox{$x \succ z$} or \mbox{$z \succ y$}.
If \mbox{$y \leq z$} or \mbox{$z \leq x$} we conclude by parts~\ref{succ>=}
or~\ref{>=succ}.
We may therefore assume that \mbox{$x > z > y$}.
Notice that 
\[
(x - y) / x = (x - z) / x + (z - y) / x < (x - z) / x + (z - y) / z.
\]
Since the left hand side is not infinitesimally close to zero,
at least one of the terms in the sum in the right hand side
must be larger than some standard number and we have shown that
the relation $\succ$ is negatively transitive.
It follows that the relation $\preceq$ is transitive.
It is reflexive by part~\ref{<}.
The relation $\sim$ is therefore reflexive, transitive and symmetric,
and part~\ref{equ} is proved.

Part~\ref{cx} is obvious.

For part~\ref{c+1}, assume \mbox{$x \succ y$}. We have \mbox{$x > y$}.
Therefore \mbox{$x + c >$} \mbox{$y + c$}.
Consider 
\[
\frac{(x + c) - (y + c)}{x + c} = \frac{x - y}{x + c} = \frac{x - y}{x}
\times \frac{x}{x + c}.
\]
If \mbox{$c / x$} is finite, so is \mbox{$(x + c) / x$} 
and \mbox{$x / (x + c)$} is not infinitesimally
close to zero.
But \mbox{$x \succ y$} implies that \mbox{$(x - y) / x$} is not
infinitesimally close to zero.
We conclude that \mbox{$x + c \succ$} \mbox{$y + c$}.

For part~\ref{c+2}, assume \mbox{$c / x$} is infinite.
Both \mbox{$\mu x / \lambda c$} and \mbox{$\mu y / \lambda c$} are
infinitesimally close to zero.
Therefore \mbox{$\lambda c \sim$} \mbox{$\mu x + \lambda c \sim$}
\mbox{$\mu y + \lambda c$}
and \mbox{$\lambda c \succ$} \mbox{$\nu x$}.

For part~\ref{negc+}, notice that \mbox{$x - c >$} \mbox{$y - c$}
and \mbox{$[(x - c) - (y - c)] / (x - c) >$} \mbox{$(x - y) / x$}.
\end{proof}
So far, the reader may get the feeling that the qualitative order
is just like the usual one. This would be a mistaken impression.
Note that \mbox{$x \preceq y$} does not imply
\mbox{$x - c \preceq$} \mbox{$y - c$}.
If $\epsilon$ is positive and infinitesimally close to zero,
\mbox{$1 + \epsilon \preceq 1$} but 
\mbox{$1 + \epsilon - ( 1 - \epsilon^{2}) \not \preceq \epsilon^{2}$}.
Note also that $\succ$ is not complete and that \mbox{$\preceq$} is not
antisymmetric.
A sequence of technical lemmas will be stated now and used
in Section~\ref{sub:char}.
Had we used the usual order $>$, they would have been obvious,
for $\prec$ they are not.
\begin{lemma}
\label{le:2}
Let \mbox{$\lambda \in ]0 , 1[$} be a standard real number.
Assume \mbox{$a , b , c \in \cR_{+}$} such that
\mbox{$a = \lambda b + (1 - \lambda) c$},
\mbox{$a \preceq b$}, \mbox{$a \preceq c$}.
Then, \mbox{$a \sim b$} and \mbox{$a \sim c$}.
\end{lemma}
\begin{proof}
Suppose, without loss of generality, that 
\mbox{$b < c$}.
Then \mbox{$b < a < c$}.
Clearly \mbox{$b \preceq a$} and \mbox{$a \sim b$}.
But \mbox{$c - a = \frac{\lambda}{1 - \lambda} (a - b)$}.
If \mbox{$\frac{a - b}{a}$} is infinitesimally close to zero,
so is \mbox{$\frac{c - a}{c}$} and \mbox{$a \sim c$}.
\end{proof}
\begin{lemma}
\label{le:n}
Let \mbox{$n > 0$}, \mbox{$\lambda_{i} > 0$}, 
\mbox{$i = 0 , \ldots , n - 1$}, standard real numbers such that
\mbox{$\sum_{i = 0}^{n - 1} \lambda_{i} = 1$}.
Let \mbox{$a , a_{i}$}, \mbox{$i = 0 , \ldots , n - 1$} be elements
of a nonstandard model $\cR_{+}$ such that
\mbox{$a = \sum_{i = 0}^{n - 1} \lambda_{i} a_{i}$} and
\mbox{$a \preceq a_{i}$} for any \mbox{$i = 0 , \ldots , n - 1$}.
Then, for any \mbox{$i = 0 , \ldots , n - 1$},
\mbox{$a \sim a_{i}$}.
\end{lemma}
\begin{proof}
By induction on $n$.
The case $n = 1$ is obvious.
The case $n = 2$ is dealt with by Lemma~\ref{le:2}.
For the general case,
consider that
\[
a = \lambda_{0} a_{0} + (1 - \lambda_{0}) \sum_{i = 1}^{n - 1} 
\frac{\lambda_{i}}{1 - \lambda_{0}} a_{i},
\]
and use Lemma~\ref{le:2} to conclude that \mbox{$a \sim a_{0}$}.
Similarly for $a_{i}$.
\end{proof}
\begin{lemma}
\label{le:basic2}
Let \mbox{$n > 0$}, \mbox{$\lambda_{i} \geq 0$}, 
\mbox{$i = 0 , \ldots , n - 1$}, standard real numbers such that
\mbox{$\sum_{i = 0}^{n - 1} \lambda_{i} = 1$}.
Let \mbox{$a , a_{i}$}, \mbox{$i = 0 , \ldots , n - 1$} be elements
of a nonstandard model $\cR_{+}$ such that
\mbox{$a = \sum_{i = 0}^{n - 1} \lambda_{i} a_{i}$} and
\mbox{$a \preceq a_{i}$} for any \mbox{$i = 0 , \ldots , n - 1$}.
Then, for any \mbox{$i = 0 , \ldots , n - 1$} such that 
\mbox{$\lambda_{i}> 0$}, 
\mbox{$a \sim a_{i}$}.
\end{lemma}
\begin{proof}
By induction on $n$.
The case $n = 1$ is obvious.
For the general case, if, for some $i$
\mbox{$\lambda_{i} = 0$}, use a straightforward induction argument.
If, for any $i$, \mbox{$\lambda_{i} > 0$}, use Lemma~\ref{le:n}.
\end{proof}

\section{Derivation of the completeness result}
We assume that the postulates {\bf A1}, {\bf A'2}, {\bf A'3}
and {\bf A''3} hold and
derive some consequences.
In most cases, the derivation is very similar to the classical case,
but we have to do with the weaker assumptions.
First, notice that, since \mbox{$p \gg q$} implies \mbox{$p > q$}
and, by {\bf A1}, the relation $>$ is asymmetric, so is $\gg$.
In particular, $\gg$ is irreflexive.
\begin{lemma}
\label{le:bet}
Assume \mbox{$p > q$}, \mbox{$\lambda \in ]0 , 1[$}. Then,
\mbox{$p > \lambda p + (1 - \lambda) q > q$}.
\end{lemma}
\begin{proof}
Assume \mbox{$p > q$}.
We know that \mbox{$p \not \gg p$}, therefore, by {\bf A'2},
for any \mbox{$\mu \in ]0 , 1[$},
\mbox{$\mu p + (1 - \mu) p > \mu q + (1 - \mu) p$}.
By taking, \mbox{$\mu = 1 - \lambda$}, 
for any \mbox{$\lambda \in ]0 , 1[$},
\mbox{$p > (1 - \lambda) q + \lambda p$}.
But we know also that \mbox{$q \not \gg p$} and therefore we have
\mbox{$\lambda p + (1 - \lambda) q > \lambda q + (1 - \lambda) q$}.
\end{proof}
The next lemma is most important.
\begin{lemma}
\label{le:mon}
Assume \mbox{$p > q$}, \mbox{$\lambda , \mu \in [0 , 1]$}, and
\mbox{$\lambda > \mu$}. Then,
\mbox{$\lambda p + (1 - \lambda) q >$}
\mbox{$\mu p + (1 - \mu) q$}.
\end{lemma}
\begin{proof}
Note that \mbox{$\mu < 1$}.
\[
\lambda p + (1 - \lambda) q = \frac{\lambda - \mu}{1 - \mu} p + 
\frac{1 - \lambda}{1 - \mu} [ \mu p + (1 - \mu) q ]
\]
By Lemma~\ref{le:bet}, \mbox{$p > \mu p + (1 - \mu) q$}.
Therefore \mbox{$\mu p + (1 - \mu) q \not \gg p$} and,
by {\bf A'2},  
\[
\frac{\lambda - \mu}{1 - \mu} p + 
\frac{1 - \lambda}{1 - \mu} [ \mu p + (1 - \mu) q ] > 
\]
\[
\frac{\lambda - \mu}{1 - \mu} [\mu p + (1 - \mu) q] +
\frac{1 - \lambda}{1 - \mu} [ \mu p + (1 - \mu) q ] =
\mu p + (1 - \mu) q.
\]
\end{proof}
\begin{corollary}
\label{co:eq1}
Let \mbox{$\lambda , \mu \in [0 , 1]$}.
If \mbox{$\lambda p + (1 - \lambda) q \sim \mu p + (1 - \mu) q$},
then either \mbox{$\lambda = \mu$} or \mbox{$p \sim q$}.
\end{corollary}
\begin{proof}
Let \mbox{$\lambda p + (1 - \lambda) q \sim \mu p + (1 - \mu) q$}.
Assume, without loss of generality, that \mbox{$p > q$}.
By lemma~\ref{le:mon}, if \mbox{$\lambda > \mu$}, we have
\mbox{$\lambda p + (1 - \lambda) q > \mu p + (1 - \mu) q$}.
We conclude that \mbox{$\mu \leq \lambda$}, and, similarly,
\mbox{$\lambda \leq \mu$}.
\end{proof}
\begin{corollary}
\label{co:unique}
For any \mbox{$p , q , r \in P$}, if \mbox{$p > q$}, there is at most one 
\mbox{$\lambda \in [0 , 1]$} such that 
\mbox{$r \sim \lambda p + (1 - \lambda) q$}.
\end{corollary}
\begin{proof}
Obvious from Corollary~\ref{co:eq1}.
\end{proof}
\begin{lemma}
\label{le:>>and>}
If \mbox{$p \gg q$} and \mbox{$q > r$},
then \mbox{$p \gg r$}.
\end{lemma}
\begin{proof}
Let \mbox{$p \gg q$} and \mbox{$q > r$}.
Let $r'$ be such that \mbox{$r > r'$}.
We shall show that, for any \mbox{$\lambda \in ]0 , 1[$},
\mbox{$\lambda r + (1 - \lambda) p \sim$}
\mbox{$\lambda r' + (1 - \lambda) p$}.
We have, \mbox{$q > r$}, \mbox{$q > r'$} and \mbox{$p \gg q$}.
Therefore 
\mbox{$\lambda q + (1 - \lambda) p \sim$}
\mbox{$\lambda r + (1 - \lambda) p$} and
\mbox{$\lambda q + (1 - \lambda) p \sim$}
\mbox{$\lambda r' + (1 - \lambda) p$}.
\end{proof}
\begin{lemma}
\label{le:exist0}
Let \mbox{$p > q > r$}.
If one assumes that there exists some \mbox{$\beta \in ]0 , 1[$}
such that
\mbox{$q >$} \mbox{$\beta p + (1 - \beta) r$}, then there
exists some \mbox{$\gamma \in ]0 , 1[$}
such that
\mbox{$q \sim$} \mbox{$\gamma p + (1 - \gamma) r$}. 
\end{lemma}
\begin{proof}
By {\bf A'3}, there is some \mbox{$\alpha \in ]0 , 1[$}
such that \mbox{$\alpha p + (1 - \alpha) r >$}
\mbox{$ q$}.
Let \mbox{$\gamma_{1}$} be the infimum of the set
\mbox{$\{\alpha \mid \alpha p + (1 - \alpha) r > q \}$}
and \mbox{$\gamma_{2}$} be the supremum of the set
\mbox{$\{\beta \mid q > \beta p + (1 - \beta) r\}$}
(by assumption this set is non-empty).
By Lemma~\ref{le:mon}, 
\mbox{$\gamma_{1} \geq \gamma_{2}$}.
Take any $\gamma$ such that 
\mbox{$\gamma_{1} \geq \gamma \geq \gamma_{2}$}.
The choice of $\gamma_{1}$ and $\gamma_{2}$, together with {\bf A1}
imply that \mbox{$q \sim \lambda p + (1 - \lambda) r$}.
\end{proof}
\begin{corollary}
\label{co:exist}
Let \mbox{$p \geq q \geq r$}, \mbox{$p > r$} and
\mbox{$p \not \gg r$}.
Then, there exists a unique \mbox{$\lambda \in [0 , 1]$}
such that
\mbox{$q \sim \lambda p + (1 - \lambda) r$}.
\end{corollary}
\begin{proof}
Uniqueness has been shown in Corollary~\ref{co:unique}.
Let us show existence.
If \mbox{$p \sim q$} or \mbox{$q \sim r$}, one can take
\mbox{$\lambda$} to be equal to one or zero respectively.
We may therefore assume that
\mbox{$p > q > r$} and \mbox{$p \not \gg r$}.
By Lemma~\ref{le:>>and>}, \mbox{$p \not \gg q$}.
By {\bf A''3} the assumptions of Lemma~\ref{le:exist0} hold
and the existence of a suitable $\lambda$ is ensured.
\end{proof}
\begin{lemma}
\label{le:both}
If \mbox{$p > p '$} and \mbox{$q > q '$}, then, for any 
\mbox{$\lambda \in [0 , 1]$}, 
\mbox{$\lambda p + (1 - \lambda) q >$}
\mbox{$\lambda p ' + (1 - \lambda) q '$}.
\end{lemma}
\begin{proof}
The cases \mbox{$ \lambda = 0$} and \mbox{$ \lambda = 1$} are easily treated.
Assume \mbox{$\lambda \in ]0 , 1[$}.
By {\bf A'2}, 
either,
\begin{enumerate}
\item \label{1}
for any \mbox{$\lambda \in ]0 , 1[$},
\mbox{$\lambda p + (1 - \lambda) q > \lambda p ' + (1 - \lambda) q$},
or 
\item \label{2}
\mbox{$q \gg p$}.
Similarly, either
\item \label{a}
for any \mbox{$\lambda \in ]0 , 1[$},
\mbox{$\lambda p ' + (1 - \lambda) q > 
\lambda p ' + (1 - \lambda) q '$},
or 
\item \label{b}
\mbox{$p ' \gg q$}.
\end{enumerate}
The only non-trivial case is when both cases~\ref{2} and~\ref{b}
occur.
But then, we have both \mbox{$q > p$} and \mbox{$p' > q$},
contradicting \mbox{$p > p'$}.
\end{proof}
\begin{lemma}
\label{le:J4}
If \mbox{$p \sim q$}, \mbox{$\lambda \in [0 , 1]$},
then \mbox{$p \sim \lambda p + (1 - \lambda) q$}.
\end{lemma}
\begin{proof}
Suppose \mbox{$p \sim q > \lambda p + (1 - \lambda) q$}.
Then, by Lemma~\ref{le:both}, 
\[
\lambda p + (1 - \lambda) q > \lambda [\lambda p + (1 - \lambda) q] 
+ (1 - \lambda) [\lambda p + (1 - \lambda) q] = \lambda p + (1 - \lambda) q,
\]
a contradiction.
Similarly \mbox{$\lambda p + (1 - \lambda) q > p \sim q$}
leads to a contradiction.
\end{proof}
\begin{lemma}
\label{le:pr>>}
Let \mbox{$\lambda \in ]0 , 1[$}.
For any \mbox{$p , q$},
\mbox{$p \not \gg \lambda q + (1 - \lambda) p$}.
\end{lemma}
\begin{proof}
Assume \mbox{$p \gg \lambda q + (1 - \lambda) p$}.
We have \mbox{$p > \lambda q + (1 - \lambda) p$}.
By Lemmas~\ref{le:bet} and~\ref{le:J4},
\mbox{$p > q$} and 
\mbox{$\lambda q + (1 - \lambda) p > q$}.
Since \mbox{$p \gg \lambda q + (1 - \lambda) p$},
we know that
\mbox{$\alpha [\lambda q + (1 - \lambda) p ] + (1 - \alpha) p \sim$}
\mbox{$\alpha q + (1 - \alpha) p$}, for any
\mbox{$\alpha \in ]0 , 1[$}.
Therefore, for any such $\alpha$,
\mbox{$\alpha \lambda q + (1 - \alpha \lambda) p \sim$}
\mbox{$\alpha q + (1 - \alpha) p$}.
By Corollary~\ref{co:eq1}, \mbox{$q \sim p$}, a contradiction.
\end{proof}
The next lemma does not mention the relation $\gg$,
but it does not hold in the wider framework in which negative utilities
are allowed:
\mbox{$1 + \epsilon \sim 1$}, but 
\mbox{$0.5 (1 + \epsilon) + 0.5 (-1) > 0.5 (1) + 0.5 (-1)$}.
Its proof is more intricate than the corresponding proof in the
classical framework, or any framework in which Independence holds.
\begin{lemma}
\label{le:J5}
Let \mbox{$\lambda \in [0 , 1]$}.
If \mbox{$p \sim q$}, then 
\mbox{$\lambda p + (1 - \lambda) r \sim$}
\mbox{$\lambda q + (1 - \lambda) r$}.
\end{lemma}
\begin{proof}
Assume \mbox{$p \sim q$}.
The result is obvious for $\lambda$ equal to one or zero.
We may assume \mbox{$\lambda \in ]0 , 1[$}.
If \mbox{$r \sim p$}, then, by Lemma~\ref{le:J4},
\mbox{$r \sim$}
\mbox{$\lambda p + (1 - \lambda) r$} and, since \mbox{$r \sim q$}
\mbox{$r \sim$}
\mbox{$\lambda q + (1 - \lambda) r$},
and the result is proved.
Assume therefore that \mbox{$p \not \sim r$}.
Without loss of generality, we may assume that
\mbox{$\lambda q + (1 - \lambda) r >$}
\mbox{$\lambda p + (1 - \lambda) r$},
We shall derive a contradiction.
Assume, first, that \mbox{$p > r$}.
Since \mbox{$p \sim q$} we have, 
\mbox{$q > r$} and, by Lemma~\ref{le:bet}, 
\mbox{$p \sim q > $}
\mbox{$\alpha q + (1 - \alpha) r > r$}, for any 
\mbox{$\alpha \in ]0 , 1[$}.
Since \mbox{$r \not \gg p$}, we have, by {\bf A'2},
for any \mbox{$\alpha \in ]0 , 1[$},
\[
\lambda p + (1 - \lambda) r >
\lambda [\alpha q + (1 - \alpha) r] + (1 - \lambda) r.
\]
But
\[
\lambda [\alpha q + (1 - \alpha) r] + (1 - \lambda) r =
\lambda \alpha q + (1 - \lambda \alpha) r =
\]
\[
\alpha [\lambda q + (1 - \lambda) r] + (1 - \alpha) r.
\]
We see that \mbox{$\lambda q + (1 - \lambda) r >$}
\mbox{$\lambda p + (1 - \lambda) r > r$}.
But
\[
\lambda p + (1 - \lambda) r >
\alpha [\lambda q + (1 - \lambda) r] + (1 - \alpha) r,
\]
for any
\mbox{$\alpha \in ]0 , 1[$},
in contradiction with {\bf A'3}.

Let us assume now that \mbox{$r > p$}.
By Lemma~\ref{le:bet}, 
\mbox{$r > $}
\mbox{$\alpha p + (1 - \alpha) r >$}
\mbox{$p \sim q$}, for any 
\mbox{$\alpha \in ]0 , 1[$}.
By Lemma~\ref{le:pr>>} \mbox{$r \not \gg \alpha p + (1 - \alpha) r$}, 
and therefore, by {\bf A'2}, we have
\mbox{$\lambda [\alpha p + (1 - \alpha) r] + (1 - \lambda) r >$}
\mbox{$\lambda q + (1 - \lambda) r$},
for any 
\mbox{$\alpha \in ]0 , 1[$}.
But we have 
\mbox{$r > \lambda q + (1 - \lambda) r >$}
\mbox{$\lambda p + (1 - \lambda) r$} and, 
\begin{equation}
\label{eq:p}
\lambda [\alpha p + (1 - \alpha) r] + (1 - \lambda) r =
\lambda \alpha p + (1 - \lambda \alpha) r =
\end{equation}
\[
\alpha [\lambda p + (1 - \lambda) r] + (1 - \alpha) r.
\]
We see that, for any
\mbox{$\alpha \in ]0 , 1[$}, 
\mbox{$\alpha [\lambda p + (1 - \lambda) r] + (1 - \alpha) r >$}
\mbox{$\lambda q + (1 - \lambda) r$}.
By {\bf A''3}, \mbox{$r \gg$} \mbox{$\lambda q + (1 - \lambda) r$},
contradicting Lemma~\ref{le:pr>>}.
\end{proof}
Our goal is now to show that the relation $\gg$ is
a weak order.
\begin{lemma}
\label{le:>>and>=}
If \mbox{$p \gg q$} and \mbox{$q \geq r$},
then \mbox{$p \gg r$}.
\end{lemma}
\begin{proof}
If \mbox{$q > r$}, the result follows from Lemma~\ref{le:>>and>}.
If \mbox{$q \sim r$}, let $r'$ be such that \mbox{$r > r'$}.
We shall show that, for any \mbox{$\lambda \in ]0 , 1[$},
\mbox{$\lambda r + (1 - \lambda) p \sim$}
\mbox{$\lambda r' + (1 - \lambda) p$}.
Indeed, by Lemma~\ref{le:J5},
\mbox{$\lambda r + (1 - \lambda) p \sim$}
\mbox{$\lambda q + (1 - \lambda) p$},
and, since \mbox{$q > r'$} and \mbox{$p \gg q$},
\mbox{$\lambda r' + (1 - \lambda) p \sim$}
\mbox{$\lambda q + (1 - \lambda) p$}.
\end{proof}
\begin{lemma}
\label{le:>and>>}
If \mbox{$p \geq q$} and \mbox{$q \gg r$},
then \mbox{$p \gg r$}.
\end{lemma}
\begin{proof}
Let \mbox{$p \geq q$} and \mbox{$q \gg r$}.
Let $r'$ be such that \mbox{$r > r'$}.
We shall show that, for any \mbox{$\lambda \in ]0 , 1[$},
\mbox{$(1 - \lambda) r + \lambda p \sim$}
\mbox{$(1 - \lambda) r' + \lambda p$}.
We know that
\mbox{$(1 - \lambda) r + \lambda q \sim$}
\mbox{$(1 - \lambda) r' + \lambda q$}.
If \mbox{$p \sim q$}, Lemma~\ref{le:J5} yields the desired conclusion.
We may therefore assume that \mbox{$p > q > r$}.
We may assume \mbox{$p \not \gg r$}.
By Lemma~\ref{co:exist}, there is some \mbox{$\alpha \in ]0 , 1[$}
such that
\mbox{$q \sim \alpha p + (1 - \alpha) r$}.
Therefore, by Lemma~\ref{le:J5},  
\mbox{$(1 - \lambda) r + \lambda q \sim$}
\mbox{$(1 - \lambda) r + \lambda [ \alpha p + (1 - \alpha) r ] =$}
\mbox{$(1 - \lambda \alpha) r + \lambda \alpha p$}.
Similarly,
\mbox{$(1 - \lambda) r' + \lambda q \sim$}
\mbox{$(1 - \lambda \alpha) r' + \lambda \alpha p$}.
We have \mbox{$r > r'$} and
\mbox{$(1 - \lambda \alpha) r + \lambda \alpha p \sim$}
\mbox{$(1 - \lambda \alpha) r' + \lambda \alpha p \sim$}.
{\bf A'2} implies \mbox{$p \gg r$}.
\end{proof}
\begin{lemma}
\label{le:negtrans}
The relation $\gg$ is negatively transitive.
\end{lemma}
\begin{proof}
Assume \mbox{$p \not \gg q$} and \mbox{$q \not \gg r$}.
We shall show that \mbox{$p \not \gg r$}.
If \mbox{$r \geq p$}, the result holds.
We may therefore assume that \mbox{$p > r$}.
If \mbox{$q \geq p$}, \mbox{$q \not \gg r$} implies
\mbox{$p \not \gg r$}, by Lemma~\ref{le:>and>>}.
We may therefore assume that \mbox{$p > q$}.
If \mbox{$r \geq q$}, \mbox{$p \not \gg q$} implies
\mbox{$p \not \gg r$}, by Lemma~\ref{le:>>and>=}.
We may therefore assume that \mbox{$q > r$}.
We have \mbox{$p > q > r$}.

\mbox{$q \not \gg r$} implies there exists some $r'$, 
\mbox{$r > r'$} and some \mbox{$\lambda \in ]0 , 1[$}, 
such that \mbox{$\lambda r + (1 - \lambda) q >$}
\mbox{$\lambda r' + (1 - \lambda) q$}.
By {\bf A'2}, since \mbox{$q \not \gg r$},
for any \mbox{$\alpha \in ]0 , 1[$}, we have
\mbox{$(1 - \alpha) r + \alpha q >$}
\mbox{$(1 - \alpha) r' + \alpha q$}.

By {\bf A''3}, since \mbox{$p \not \gg q$}, there is some
\mbox{$\gamma \in ]0 , 1[$} such that
\mbox{$q > \gamma p + (1 - \gamma) r$}.
By Lemma~\ref{le:exist0}, there is some
\mbox{$\beta \in ]0 , 1[$} such that
\mbox{$q \sim \beta p + (1 - \beta) r$}.
By Lemma~\ref{le:J5},
\[
(1 - \alpha) r + \alpha [ \beta p + (1 - \beta) r ] >
(1 - \alpha) r' + \alpha [ \beta p + (1 - \beta) r ] 
\]
and
\[
\alpha \beta p + (1 - \alpha \beta) r >
\alpha \beta p + (1 - \alpha \beta) r' 
\]
and we conclude that \mbox{$p \not \gg r$}.
\end{proof}
Let \mbox{$p \asymp q$} denote that
\mbox{$p \not \gg q$} and \mbox{$q \not \gg p$}.
\begin{corollary}
\label{co:weak}
The relation $\gg$ is a weak order and therefore the relation $\asymp$
is an equivalence relation.
\end{corollary}
\begin{proof}
We notice that $\gg$ is asymmetric, by {\bf A1}.
Lemma~\ref{le:negtrans} shows that it is negatively transitive.
\end{proof}
\begin{lemma}
\label{le:vNM}
Let \mbox{$A \subseteq P$} be an equivalence class
of $P$ under $\asymp$.
The relation $>$ on the set $A$ satisfies
{\bf A1}, {\bf A2} and {\bf A3}.
It may therefore be represented by a linear functional
\mbox{$u : A \longrightarrow {\bf R}$}, the standard set of real numbers.
If the set $P$ is finitely generated, then, the functional $u$
may be taken into the positive real numbers.
\end{lemma}
\begin{proof}
Since $>$ is a weak order on $P$, it is a weak order on 
\mbox{$A \subseteq P$}.
For any \mbox{$p , q \in A$},
\mbox{$p \not \gg q$}, therefore {\bf A'2} implies
{\bf A2} and {\bf A'3} and {\bf A''3} imply 
{\bf A3}.
By Theorem 4.1 of~\cite{Fish:handbook}, the relation $>$ on $A$
may be represented by a linear functional on $A$.
This functional is uniquely defined up to a positive affine transformation.
If the set $P$ is finitely generated, by Lemma~\ref{le:mon},
the range of $u$ is bounded and therefore the addition of a large
enough constant to $u$ ensures its range contains only positive numbers.
\end{proof}
\begin{lemma}
\label{le:fin}
For any \mbox{$p , q \in P$}, \mbox{$\lambda \in [0 , 1]$}, one of the following two cases obtains:
\begin{itemize}
\item
\mbox{$\lambda p + (1 - \lambda) q \asymp p$}, or 
\item
\mbox{$\lambda p + (1 - \lambda) q \asymp q$}.
\end{itemize}
\end{lemma}
\begin{proof}
If $\lambda$ is equal to zero or to one, the result is obvious.
Assume \mbox{$\lambda \in ]0 , 1[$}.
If \mbox{$p \sim q$}, Lemma~\ref{le:J4} implies 
\mbox{$\lambda p + (1 - \lambda) q \sim p$} and therefore 
\mbox{$\lambda p + (1 - \lambda) q \asymp p$}.
Without loss of generality, assume \mbox{$p > q$}.
Then, by Lemma~\ref{le:bet}, 
\mbox{$p > \lambda p + (1 - \lambda) q$} and therefore
\mbox{$\lambda p + (1 - \lambda) q \not \gg p$}.
By Lemma~\ref{le:pr>>}, 
\mbox{$p \not \gg \lambda q + (1 - \lambda) p$}.
\end{proof}
The proof of Theorem~\ref{the:main} will now be presented.
\setcounter{theorem}{1}
\addtocounter{theorem}{-1}
\begin{theorem}
If the set $P$ is finitely generated, then the relation
$>$ satisfies {\bf A1}, {\bf A'2}, {\bf A'3} and {\bf A''3}
iff there is some nonstandard model \cR\ and some
linear functional \mbox{$u : P \longrightarrow \cR_{+}$} such that
\mbox{$\forall p , q \in P$}, \mbox{$p > q$} iff
\mbox{$u(p) \succ u(q)$}.
\end{theorem}
\begin{proof}
The {\em if} part has been proven in Section~\ref{sec:post}.
Let us deal with the {\em only if} part.
By Lemma~\ref{le:fin} any element of $P$ is 
$\asymp$-equivalent to one of the generators.
Therefore there is a finite number of $\asymp$-equivalence
classes.
Let them be ordered by $\gg$:
\mbox{$A_{0} \gg A_{1} \gg \ldots A_{n - 1}$}.
Let us choose any non-standard model \cR and
any \mbox{$\epsilon \in \cR$}, 
positive and infinitesimally close to zero.
By Lemma~\ref{le:vNM}, there are linear functions:
\mbox{$u_{i} : A_{i} \longrightarrow {\bf R}$} such that
\mbox{$\forall p , q \in A_{i}, p > q$} iff 
\mbox{$u_{i}(p) > u_{i}(q)$}.
We shall define, for every \mbox{$ p \in P$}: 
\mbox{$u(p) \eqdef \epsilon^{i} u_{i}(p)$}.

We shall show that, if \mbox{$p > q$}, then 
\mbox{$u(p) \succ u(q)$}.
Suppose \mbox{$p > q$}.
If \mbox{$p \gg q$}, \mbox{$u(p) / u(q)$} is infinite
and \mbox{$u(p) \succ u(q)$}.
If \mbox{$p \not \gg q$}, then \mbox{$p \asymp q$}.
Suppose \mbox{$p , q \in A_{i}$}.
Since $u_{i}(p)$ and $u_{i}(q)$ are standard numbers such
that \mbox{$u_{i}(p) > u_{i}(q)$}, we have
\mbox{$u_{i}(p) \succ u_{i}(q)$} and 
\mbox{$\epsilon^{i} u_{i}(p) \succ \epsilon^{i} u_{i}(q)$}.

We shall show now that, if \mbox{$u(p) \succ u(q)$}, then
\mbox{$p > q$}.
Assume \mbox{$u(p) \succ u(q)$}.
If \mbox{$p \asymp q$}, then 
\mbox{$\epsilon^{i} u_{i}(p) \succ \epsilon^{i} u_{i}(q)$}
and therefore \mbox{$u_{i}(p) > u_{i}(q)$} and
\mbox{$p > q$}.
If \mbox{$p \not \asymp q$}, then there are $i , j$,
\mbox{$i < j$} such that \mbox{$p \in A_{i}$} and
\mbox{$q \in A_{j}$} and therefore \mbox{$p \gg q$}
and \mbox{$p > q$}.
\end{proof}

\section{Proof of Theorem~\ref{the:subj}}
\label{appen:subj}
\setcounter{theorem}{2}
\addtocounter{theorem}{-1}
\begin{theorem}
Assume $P$ is finitely generated and $>$ is a binary relation on ${\bf F}$.
There is a probability measure $p$ on $S$, a nonstandard model $\cR$ 
and a linear function \mbox{$u : P \longrightarrow \cR_{+}$} such that
\mbox{$\forall a , b \in {\bf F}$}, 
\begin{equation}
\label{eq:char}
a > b \Leftrightarrow
\sum_{s \in S} p(s) u(a(s)) \succ \sum_{s \in S} p(s) u(b(s))
\end{equation}
iff the relation $>$ satisfies {\bf A1}, {\bf A'2}, {\bf A'3}, 
{\bf A''3}, {\bf A'4} and {\bf A'5}.
If the relation $>$ on $P$ is not trivial, 
the probability measure $p$ is unique.
\end{theorem}
\begin{proof}
Let us prove first the {\em only if} part.
Assume Equation~\ref{eq:char} holds.
We notice that ${\bf F}$ is a mixture set and that,
by letting \mbox{$u'(a) \eqdef \sum_{s \in S} p(s) u(a(s))$},
we have 
\[
a > b \Leftrightarrow u'(a) \succ u'(b).
\]
By the results of Section~\ref{sec:post},
{\bf A1}, {\bf A'2}, {\bf A'3}, 
and {\bf A''3} hold.

Let us check that {\bf A'4} holds. Assume that \mbox{$a(s) = b(s)$} for any
\mbox{$s \in S$}, \mbox{$s \neq s_{0}$} and that \mbox{$a > b$}.
If 
\[
\sum_{s \in S} p(s) u(a(s)) \succ \sum_{s \in S} p(s) u(b(s)),
\]
then, by Lemma~\ref{le:basic}, part~\ref{negc+},
\mbox{$p(s_{0}) u(a(s_{0})) \succ p(s_{0}) u(b(s_{0}))$}.
By part~\ref{cx}, we have
\mbox{$u(a(s_{0})) \succ u(b(s_{0}))$}.
But \mbox{$\sum_{s \in S} p(s) u(a(s_{0})) = u(a(s_{0}))$}
and similarly
\mbox{$\sum_{s \in S} p(s) u(b(s_{0})) = u(b(s_{0}))$}.
Therefore
\mbox{$\sum_{s \in S} p(s) u(a(s_{0})) \succ \sum_{s \in S} p(s) u(b(s_{0}))$}
and \mbox{$a(s_{0}) > b(s_{0})$}.

Let us check now that {\bf A'5} holds.
By Lemma~\ref{le:inf}, \mbox{$a(t) \gg a$} implies that either 
\mbox{$u(a(t)) / u(a)$} is infinite, or \mbox{$a(t) > a$}
and \mbox{$a \leq b$} for any 
\mbox{$b \in {\bf F}$}.
Since \mbox{$u(a) \geq p(t) u(a(t))$}, if
\mbox{$u(a(t)) / u(a)$} is infinite,
\mbox{$p(t)$} must be equal to zero and
$t$ is null.
If \mbox{$a(t) > a$} and \mbox{$a \leq b$} for any 
\mbox{$b \in {\bf F}$}, then, letting
\mbox{$x = \sum_{s \in S} p(s) u(a(s))$}, we have
\mbox{$x \preceq u(a(s))$} for any \mbox{$s \in S$}.
By Lemma~\ref{le:basic2}, \mbox{$x \not \sim u(a(t))$} implies
that \mbox{$p(t) = 0$}, and $t$ is null.

Let us deal now with the more challenging {\em if} part.
Assume {\bf A1}, {\bf A'2}, {\bf A'3}, 
{\bf A''3}, {\bf A'4} and {\bf A'5} are satisfied.
By Theorem~\ref{the:main}, since {\bf A1}, {\bf A'2},
{\bf A'3} and {\bf A''3} hold, there is a nonstandard model $\cR$
and a linear function
\mbox{$u^{\star} : {\bf F} \longrightarrow \cR_{+}$} such that
\mbox{$\forall a , b \in {\bf F}$}, 
\begin{equation}
\label{eq:star}
a > b \Leftrightarrow
u^{\star}(a) \succ u^{\star}(b).
\end{equation}
Since $P$ is finitely generated, by Lemma~\ref{le:bet},
there are minimal ($l$) and maximal ($h$) elements of $P$. 
Both $l$ and $h$ are constant functions in ${\bf F}$.
By Lemma~\ref{le:A'4}, for any \mbox{$a \in {\bf F}$},
\mbox{$l \preceq a \preceq h$}.
If \mbox{$l \sim h$}, all acts are equivalent and the Theorem
is easily proved.
Suppose, therefore, that \mbox{$h \succ l$}.
By Lemma~\ref{le:basic}, parts~\ref{cx} and~\ref{negc+},
\[
a > b \Leftrightarrow \frac{u^{\star}(a) - u^{\star}(l)}{u^{\star}(h)}
\succ \frac{u^{\star}(b) - u^{\star}(l)}{u^{\star}(h)}.
\]
This means we may assume, from now on, that \mbox{$u^{\star}(l) = 0$}
and \mbox{$u^{\star}(h) = 1$}.
For \mbox{$t \in S$}, let us define \mbox{$c_{t} \in {\bf F}$}
by \mbox{$c_{t}(t) = h$} and, for \mbox{$s \neq t$},
\mbox{$c_{t}(s) = l$}.
If $t$ is not null, by {\bf A'5}, \mbox{$h \not \gg c_{t}$}
and, by {\bf A''3} and Lemma~\ref{le:exist0} there is some 
\mbox{$\lambda \in [0 , 1]$} such that
\mbox{$c_{t} \sim \lambda h + (1 - \lambda) l$}; in this case we shall
define \mbox{$p(t) = \lambda$} and notice that 
\mbox{$u^{\star}(c_{t}) \sim p(t)$}.
If $t$ is null we take
\mbox{$p(t) = 0$}.
In this case, notice that 
\mbox{$c_{t} \sim l$} and therefore
\mbox{$u^{\star}(c_{t}) = 0$}.

Clearly, for any \mbox{$s \in S$},
\mbox{$0 \leq p(s) \leq 1$}.
We claim that, for any \mbox{$a \in {\bf F}$},
\[
u^{\star}(a) \sim \sum_{s \in S} p(s) \: u^{\star}(a(s)).
\]
Notice that we do not claim equality, only {\em similarity}, i.e.,
$\sim$, between the left hand side and the right hand side.
Let us justify our claim now.
Let \mbox{$a \in {\bf F}$} and
\mbox{$c = \sum_{s \in S} u^{\star}(a(s))$}.
If \mbox{$c = 0$}, then, for any \mbox{$s \in S$},
\mbox{$u^{\star}(a(s)) = 0$} and
\mbox{$a(s) \sim l$}, and by Lemma~\ref{le:A'4},
\mbox{$a \sim l$}, \mbox{$u^{\star}(a) = 0$},
by Lemma~\ref{le:basic}, part~\ref{zero} and
the result holds.
If \mbox{$c > 0$}, let
\[
b \eqdef \sum_{s \in S} \frac{u^{\star}(a(s))}{c} \: c_{s}.
\]
Let us, first, compare $a$ and $b$. 
For any \mbox{$t \in S$}, 
\[
b(t) = \frac{u^{\star}(a(t))}{c} \: h + (1 - \frac{u^{\star}(a(t))}{c}) \: l.
\]
We see that, for any \mbox{$t \in S$},
\begin{equation}
\label{eq:t}
u^{\star}(b(t)) = \frac{1}{c} \: u^{\star}(a(t)).
\end{equation}
Let us assume, first, that
\mbox{$c \leq 1$}.
Define the act $a'$ by:
\[
a' \eqdef c \: b + (1 - c) \: l.
\]
For any \mbox{$s \in S$}, 
\mbox{$a'(s) = c \: b(s) + (1 - c) l$},
\mbox{$u^{\star}(a'(s)) = c \: u^{\star}(b(s))$} and therefore
\mbox{$u^{\star}(a'(s)) =  u^{\star}(a(s))$}.
We conclude, by Lemma~\ref{le:A'4}, that \mbox{$a' \sim a$},
and therefore \mbox{$u^{\star}(a) \sim  u^{\star}(a')$}.
But, by linearity of $u^{\star}$, 
\mbox{$u^{\star}(a') = c \: u^{\star}(b)$}.
We conclude that \mbox{$u^{\star}(a) \sim c \: u^{\star}(b)$}.
The case \mbox{$c > 1$} is treated similarly by defining
\[
b' \eqdef (1 / c) \: a + (1 - 1 / c) \: l.
\]
But, now, by linearity of $u^{\star}$,
\[
u^{\star}(b) = \frac{1}{c} \: \sum_{s \in S} u^{\star}(a(s)) \: 
u^{\star}(c_{s}).
\]
Therefore
\[
u^{\star}(a) \sim \sum_{s \in S} u^{\star}(a(s)) \: u^{\star}(c_{s}).
\]
We have seen that \mbox{$u^{\star}(c_{s}) \sim p(s)$} when \mbox{$p(s) > 0$}
and \mbox{$u^{\star}(c_{s}) = p(s) $} when \mbox{$p(s) = 0$}.
Therefore
\[
u^{\star}(a) \sim \sum_{s \in S} p(s) \: u^{\star}(a(s)),
\]
as claimed.
The first conclusion we draw is that:
\[
a > b \Leftrightarrow
\sum_{s \in S} p(s) \: u^{\star}(a(s)) \succ 
\sum_{s \in S} p(s) \: u^{\star}(b(s)).
\]
Our second conclusion is that
\mbox{$1 =$} \mbox{$u^{\star}(h) \sim$}
\mbox{$\sum_{s \in S} p(s) u^{\star}(h) =$}
\mbox{$\sum_{s \in S} p(s)$}.
Therefore \mbox{$\sum_{s \in S} p(s) \sim 1$}.
But, for any \mbox{$s \in S$}, $p(s)$ is {\em standard}.
It follows that \mbox{$\sum_{s \in S} p(s) = 1$}.

For the uniqueness claim, notice that
\mbox{$\sum_{s \in S} p(s) u(c_{t}(s)) = p(t) u(h) + (1 - p(t)) u(l)$}.
By Equation~\ref{eq:char}, then, \mbox{$c_{t} \sim p(t) h + (1 - p(t)) l$}.
By Corollary~\ref{co:exist}, there is only one such $p(t)$.
\end{proof}



\section*{Acknowledgment}
I want to thank Israel Aumann for his encouragements
and suggestions to improve the presentation of this work
and Aviad Heifetz who helped correct an argument.
An anonymous referee provided unusually helpful insights, two of which
have been directly incorporated into the paper, thanks a lot!

\bibliographystyle{plain}

\end{document}